\documentclass[journal,twoside,a4paper]{IEEEtran}
\usepackage[brazil]{babel}
\usepackage[utf8]{inputenc}
\usepackage[]{graphicx}
   
\begin{document}

\bibliographystyle{IEEEtran}

\title{\LARGE O Algoritmo Usado No Programa de Criptografia PASME}

\author{Péricles Lopes Machado
\thanks{Laboratório de Análises Numéricas em Eletromagnetismo (LANE), Universidade Federal do Pará,
         caixa postal 8619, CEP 66073-900, Brasil; e-mail: pericles.machado@itec.ufpa.br
       }
}

\maketitle

\begin{abstractEng}
This work will present the main encryption algorithm  of the PASME tool,
PASME  allows encrypt and hide information in various types of files.
The algorithm uses the fact that factoring large numbers is a difficult issue in terms of computational
performing to make the main steps of the encryption.
\end{abstractEng}

\begin{abstract}

Neste trabalho será apresentado o principal algoritmo de criptografia da ferramenta PASME,
 a qual permite encriptação e ocultamento de informações em diversos tipos de arquivos. 
O algoritmo utiliza o fato da fatoração de números grandes ser um problema difícil do ponto de vista computacional, 
efetuando assim, os principais passos da encriptação. 

\end{abstract}

\IEEEoverridecommandlockouts

\begin{keywords}
Criptografia, Teoria dos números, Teoria da informação
\end{keywords}

\section{Introdução}

A ideia fundamental de qualquer algoritmo de criptografia é modificar a representação de uma informação para garantir
proteção contra acesso indevidos.

No decorrer dos anos, muitos algoritmos de criptografia foram desenvolvidos. Um dos mais antigos realiza uma permutação no 
alfabeto que contém todos os símbolos da mensagem que será encriptada. Contudo, este algoritmo apresenta grande vulnerabilidade 
a uma análise da frequência de ocorrência de determinados símbolos, principalmente quando aplicados à textos escritos. 

Outro método clássico, usado em mensagens binárias, consiste em inverter certos bits e armazenar a posição dos bits 
que foram invertidos em outra palavra, a folha-chave, com o mesmo tamanho da mensagem que foi encriptada. 
Um problema desse método é que o tamanho da folha-chave pode ser muito grande, inviabilizando o processo de encriptação.

Muitos algoritmos modernos utilizam estratégias envolvendo teoria dos números através 
da utilização de problemas que atualmente são intratáveis do ponto de vista computacional.
Um exemplo clássico desta classe de algoritmo é o RSA ~\cite{book.algoritmos} ~\cite{book.matematicadiscreta}.

A ideia do presente trabalho é utilizar a intratabilidade da fatoração de inteiros grandes para realizar
 os passos-chave de sua encriptação. Nas próximas seções, serão descritos os passos realizados pelo algoritmo de encriptação PASME, 
além de serem comentados alguns detalhes de sua implementação ~\cite{site.pasme}.

\section{Algumas funções fundamentais}

\subsection{A função $\mp$ (inflar)}

A ideia fundamental do algoritmo PASME é a mudança na base de representação de um número. Mudar
a base de representação de um número inteiro $n=a_0a_1a_2...a_k$ para a base $b$
consiste em realizar a operação descrita na equação \ref{eq.Mudanca.Base}

\begin{equation}
T(n,b)=a_{0}b^{k}+a_{1}b^{k-1}+...+a_{k}b^{0}
\label{eq.Mudanca.Base}
\end{equation} 

A função $\mp$ descrita em \ref{eq.mp} é uma mudança de base onde a cada digito é adicionado um "lixo".

\begin{equation}
\mp(n,b,v)=(a_{0}+c_0)b^{1}+(a_{1}+c_{1})b^{2}+...+(a_{k}+c_{k})b^{k+1}
\label{eq.mp}
\end{equation}

Onde $c_{i}$ é descrito na equação \ref{eq.mp.ci}.

\begin{eqnarray}
c_{i}=
\left 
\{ \begin{array}{clcr}
	\triangleright(v) & ,se & i=0\\
	\triangleright(c_{i-1}) & ,se & i>0
	\end{array}
\right. 
\label{eq.mp.ci}
\end{eqnarray}

Nas equações \ref{eq.mp} e \ref{eq.mp.ci}, $\triangleright(x)$ é o próximo primo depois de $x$, 
$v$ é um inteiro qualquer, $n$ é a informação representada na forma de um inteiro, $a_{k}$ é um digito de $n$ na base original 
e $b$ é a base alvo. 

\subsection{A função $\pm$ (sujar)} 

$\pm$ é semelhante a função $\mp$, só que o "lixo" $v$ usado é o mesmo em todos dígitos, conforme pode ser visto na equação \ref{eq.pm}.

\begin{equation}
\pm(n,b,v) = (a_{0}+v)b^{0}+(a_{1}+v)b^{1}+...+(a_{k}+v)b^{k}
\label{eq.pm}
\end{equation}

\section{O algoritmo de encriptação PASME \label{desc.pasme.alg} }

A seguir, serão descritos os procedimentos para encriptar ou desencriptar uma mensagem usando o algoritmo PASME.
O algoritmo $PASME(n,key)$ encripta uma mensagem $n$ usando a frase-chave $key$.

\subsection{Encriptando uma mensagem}

O processo de encriptação inicia com a geração de 7 números aleatórios 
(de preferência, grandes) $r_i, i=1...7$. Em seguida, são definidos 4 números 
$K_i=\triangleright(r_i)$, para $i=1...5$ e $i\neq3$, $K_3=\triangleright(K_5+d_{max}+r_3+1)$, $d_{max}$ é o maior digito da base em que a informação
originalmente está representada.  

Para continuar  o processo de encriptação, uma frase-chave $key$ tem de ser fornecida. Usando-se a frase-chave, são gerados
os números $W=\mp(key,K_{3},K_{2})+K_{1}$, $Q=\triangleright(\pm(n,K_3,K_5)+r_7)$, $P=WQ+K_4$, 
 e $X=\pm(n,K_3,K_5)$ xor $Q$. $X$ é a mensagem $n$ encriptada.

As informações divulgadas são os números $K_i(i=1...5)$, $P$ e $X$.

\subsection{Desencriptando uma mensagem }

Para desencriptar uma mensagem, é preciso que sejam fornecidos os números $K_i(i=1...5)$, $P$ e $X$, além da frase-chave $key$.

O primeiro passo da desencriptação é a validação da chave, para realizar essa operação, gera-se o número $W'=\mp(key,K_{3},K_{2})+K_{1}$ 
e é verificado se $P$ mod $W' = K_4$. Efetuada a validação, pode-se recuperar $Q=(P-K_4)/W'$ . 

Com $Q$ recuperado, a mensagem $n$ 
ocultada em $X$ poderá ser revelada. Para revelar a mensagem $n$, gera-se o número $Y=X$ xor $Q$ e 
o procedimento descrito a seguir tem de ser efetuado.

\begin{enumerate}
	\item $X'=\emptyset$, $X'$ é uma palavra vazia
	\item Enquanto Y $\neq$ 0:
	\begin{enumerate}
		\item $a\leftarrow Y$ mod $K_3$ 
		\item $Y\leftarrow Y-a$
		\item $Y\leftarrow Y/K_3$, efetua-se a divisão inteira de $Y$ por $K_3$.
		\item $a\leftarrow a-K_5$
		\item $X'\leftarrow X'\oplus a$, $\oplus$ é a operação de concatenação, ou seja, a união de duas palavras
(por exemplo,$33\oplus 5 = 335$).
	\end{enumerate}
	\item $X'$ é a mensagem desencriptada
\end{enumerate}

\section{Comentários sobre a implementação de PASME disponível em ~\cite{site.pasme} }

Em ~\cite{site.pasme} está disponível uma implementação do algoritmo de criptografia descrito na seção \ref{desc.pasme.alg}.
Essa implementação utiliza a biblioteca GMP ~\cite{site.gmp} para realizar as operações envolvendo inteiros presentes no algoritmo PASME. Como
a ferramenta ~\cite{site.pasme}  permite encriptar arquivos com tamanho variáveis, usar o algoritmo PASME nem sempre é uma boa escolha, 
já que dependendo do tamanho da mensagem o tempo de execução pode ser alto. Então, por questões de eficiência, a implementação ~\cite{site.pasme}
utiliza o processo de encriptação de dois passos descrito a seguir para encriptar uma mensagem $n$. 

\begin{enumerate}
	\item Gera-se uma folha-chave $fc$ com um tamanho de $L(fc)$ bytes.
	\item Cria-se aleatoriamente uma frase-chave $key$ com $L(key)$ bytes de tamanho.
	\item Utiliza-se o algoritmo descrito em \ref{desc.pasme.alg} para encriptar a folha-chave $fc$.
	\item Quebra-se a mensagem $n$ em $L(n)$ bytes,
	\item $i\leftarrow 0$
	\item $k\leftarrow 0$
	\item $X\leftarrow \emptyset$
	\item Enquanto $i\leq L(n)$:
	\begin{enumerate}
		\item $X\leftarrow X \oplus (n_i$ xor $fc_k)$, $n_i$ é i-ésimo byte da mensagem $n$ e $fc_k$ é o k-ésimo byte da folha-chave $fc$.
		\item $i\leftarrow i+1$
		\item $k\leftarrow (k+1)$ mod $L(fc)$
	\end{enumerate}
\end{enumerate}

Para desencriptar, o passo (1) do algoritmo anterior não é executado,
no passo (2)  é fornecido a frase-chave que "abre" a mensagem e  no passo (3) é chamado o algoritmo de desencriptação descrito em \ref{desc.pasme.alg}. 

Na implementação ~\cite{site.pasme},
cada simbolo (digito num número) tem 1 byte (8 bits) de comprimento.

A implementação ~\cite{site.pasme} armazena em um arquivo-alvo as informações públicas geradas pelo algoritmo \ref{desc.pasme.alg} e a 
mensagem $X$ gerada pelo procedimento anterior. 

Para ocultar informações em arquivos, ~\cite{site.pasme} primeiramente verifica o tamanho, em
bytes, da informação que será ocultada. Após isso, a informação é concatenada ao arquivo e, por fim, concatena-se o tamanho
da informação (em ~\cite{site.pasme}, um inteiro com 4 bytes de comprimento). O procedimento para recuperar a 
informação é semelhante, só que, primeiramente, recupera-se o tamanho $L$ (em ~\cite{site.pasme}, os 4 últimos bytes do arquivo) da informação que está oculta,
depois recua-se $L-4$ bytes a partir do fim do arquivo, no caso de ~\cite{site.pasme}, e armazena-se os $L$ bytes seguintes em um arquivo-alvo.

A interface gráfica da implementação ~\cite{site.pasme} foi criada utilizando-se o framework QT4 ~\cite{site.qt}.

\section{Conclusões}
Este trabalho apresentou um algoritmo de encriptação que usa o fato da mesma informação ter
significados distintos dependendo da base em que está representada e de, atualmente, certos
problemas em teoria dos números serem intratáveis. Tal algoritmo faz parte da ferramenta PASME
que permite a encriptação e ocultamento da informação em arquivos nos mais diversos formatos.

\section{Agradecimentos}
O autor agradece a Diego Aranha por apontar uma falha no algoritmo inicial, a João Augusto Palmitesta Neto por sugestões e testes na
implemtentação ~\cite{site.pasme} do algoritmo e Fabio Lobato por revisar o artigo.

\bibliography{artigo}
\end{document}